\documentclass[twocolumn,showpacs,preprintnumbers,amsmath,amssymb]{revtex4}
\usepackage{graphicx,bm,color}

\def\be{\begin{equation}}
\def\ee{\end{equation}}
\def\bea{\begin{eqnarray}}
\def\eea{\end{eqnarray}}
\def\nnb{\nonumber}
\def\bbuildrel#1_#2^#3{\mathrel{\mathop{\kern 0pt#1}\limits_{#2}^{#3}}}

\newcommand{\scs}{\scriptscriptstyle}
\newcommand{\f}{\frac}
\newcommand{\fm}[2]{{\textstyle \frac{#1}{#2}}}

\begin{document}
\preprint{IFT-13/2010, TTP10-40, SFB/CPP-10-87}
\title{Completing the Calculation of BLM corrections to {\boldmath $\bar B \to X_s \gamma$}}
\author{Miko{\l}aj Misiak and Micha{\l} Poradzi\'nski}
\affiliation{Institute of Theoretical Physics, University of Warsaw, PL-00-681 Warsaw, Poland}
\affiliation{Institut f\"ur Theoretische Teilchenphysik, Karlsruhe Institute of Technology,
          D-76128 Karlsruhe, Germany}
%
\begin{abstract}
  Perturbative ${\cal O}(\alpha_s^2)$ corrections to ${\cal B}(\bar B \to X_s
  \gamma)$ in the BLM approximation receive contributions from two-, three-
  and four-body final states. While all the two-body results are well
  established by now, the other ones have remained incomplete for several
  years. Here, we calculate the last contribution that has been missing to
  date, namely the one originating from interference of the current-current
  and gluonic dipole operators ($K_{18}^{(2)\beta_0}$ and
  $K_{28}^{(2)\beta_0}$).  Moreover, we confirm all the previously known
  results for BLM corrections to the photon energy spectrum that involve the
  current-current operators (e.g., $K_{22}^{(2)\beta_0}$ and
  $K_{27}^{(2)\beta_0}$). Finally, we also confirm the recent findings of
  Ferroglia and Haisch on self-interference of the gluonic dipole operator
  ($K_{88}^{(2)\beta_0}$).
\end{abstract}
\maketitle
\section{Introduction \label{sec:intro}}

Weak radiative decay of the $B$ meson is a well-known probe of physics beyond
the Standard Model (SM). Calculations of its inclusive branching ratio in the
SM for $E_\gamma > 1.6\;$GeV give~\cite{Misiak:2006zs,Misiak:2006ab}
\be \label{bsgsm}
{\cal B}(\bar B \to X_s \gamma)_{\scs\rm SM} = \left( 3.15 \pm 0.23 \right)\times 10^{-4},
\ee
which agrees within $1.2\sigma$ with the world average~\cite{Barberio:2008fa}
\be \label{bsgexp}
{\cal B}(\bar B \to X_s \gamma)_{\rm exp} = \left( 3.55 \pm 0.24 \pm 0.09 \right)\times 10^{-4}.
\ee
The above experimental result includes a model uncertainty that is due to
averaging several measurements with various photon energy cuts $E_0$ and
extrapolating them to $E_0 = 1.6\;$GeV where the theory prediction is most
reliable. Measurements with energy cuts $1.8\;{\rm GeV} \leq E_0 \leq
2.0\;{\rm GeV}$~\cite{Limosani:2009qg,Aubert:2006gg,Chen:2001fj} have
significantly smaller background-subtraction errors than those with $E_0 =
1.7\;$GeV~\cite{Limosani:2009qg}. More work at balancing model-dependence and
background-subtraction uncertainties is necessary in the future to obtain
accurate experimental averages.

As far as the SM calculations are concerned, further improvements require
another critical re-analysis of non-perturbative effects~\cite{Benzke:2010js},
as well as a full perturbative ${\cal O}(\alpha_s^2)$ evaluation of $\Gamma(b
\to X_s^p \gamma)$, where $X_s^p$ stands for $s$, $sg$ and $sq\bar q$ partonic
states ($q=u,d,s$). Such calculations are most conveniently performed in the
framework of an effective low-energy theory that arises from the SM via
decoupling of the $W$ boson and all the heavier particles. So long as
higher-order electroweak and/or CKM-suppressed effects are neglected, the
relevant flavor-changing weak interactions at the 
renormalization scale~ $\mu_b \sim m_b/2$~ are given by
\be
{\cal L}_{\rm weak} = \f{4G_F}{\sqrt{2}} \sum_{i=1}^8 C_i(\mu_b) Q_i,
\ee
where $Q_i$ denote either dipole-type or four-quark operators (see below),
and $C_i(\mu_b)$ stand for their Wilson coefficients.

Following Refs.~\cite{Misiak:2006ab,Gambino:2001ew}, we shall normalize the
radiative decay rate to the charmless semileptonic one, and parametrize their
rato in terms of symmetric matrices $K_{ij}(\mu_b,E_0)$ as follows:
\be \label{pert.ratio}
\f{\Gamma[ b \to X_s \gamma]}{\Gamma[ b \to X_u e \bar{\nu}]} = 
\left| \f{ V^*_{ts} V_{tb}}{V_{ub}} \right|^2 
\f{6 \alpha_{\rm em}}{\pi} \sum_{i,j=1}^8 C_i^{\rm eff} C_j^{\rm eff} K_{ij},
\ee
where $C_i^{\rm eff}$ are certain linear combinations of $C_i$, see Eq.~(5) of
Ref.~\cite{Chetyrkin:1996vx}. Evaluation of all the $C_i^{\rm eff}(\mu_b)$ up
to the Next-to-Next-to-Leading Order (NNLO) in QCD has been already completed
several years ago~\cite{Bobeth:1999mk}.

In the perturbative expansion of $K_{ij}$ 
\be \label{kexp}
K_{ij} = K_{ij}^{(0)} + \f{\alpha_s}{4\pi} K_{ij}^{(1)} 
+ \left( \f{\alpha_s}{4\pi}\right)^2 K_{ij}^{(2)} + \ldots
\ee
all the ${\cal O}(1)$ and ${\cal O}(\alpha_s)$ terms are known since a long
time~\cite{Buras:2002er}. As far as $K_{ij}^{(2)}$ are concerned, the
so-called penguin four-quark operators $Q_3$, \ldots, $Q_6$ can be neglected
thanks to smallness of their Wilson coefficients. We can restrict our
attention to
\bea
Q_1 &=& (\bar{s}_L \gamma_{\mu} T^a c_L) (\bar{c}_L \gamma^{\mu} T^a b_L),\nnb\\[1mm]
Q_2 &=& (\bar{s}_L \gamma_{\mu}     c_L) (\bar{c}_L \gamma^{\mu}     b_L),\nnb\\[1mm]
Q_7 &=&  \f{e}{16\pi^2} m_b (\bar{s}_L \sigma^{\mu \nu}     b_R) F_{\mu \nu},\nnb\\[1mm]
Q_8 &=&  \f{g}{16\pi^2} m_b (\bar{s}_L \sigma^{\mu \nu} T^a b_R) G_{\mu \nu}^a,
\eea
i.e. consider $K_{ij}^{(2)}$ with $i,j \in \{1,2,7,8\}$ only.

At present, $K_{ij}^{(2)}$ are known in a complete
manner~\cite{Melnikov:2005bx,Asatrian:2010rq,Asatrian:2006rq} for $(ij)=(77)$
and $(78)$, while the other cases are
estimated~\cite{Ligeti:1999ea,Bieri:2003ue,Ferroglia:2010xe} using the
BLM~\cite{Brodsky:1982gc} approximation. In Ref.~\cite{Misiak:2006ab}, non-BLM
contributions to the decay rate have been calculated in the $m_c \gg m_b/2$
limit, and then interpolated downwards in $m_c$ assuming that they
vanish at $m_c=0$. Such a treatment of non-BLM NNLO corrections in the
evaluation of Eq.~(\ref{bsgsm}) still remains the current state-of-art for the
numerically important but yet unknown $K_{17}^{(2)}$ and $K_{27}^{(2)}$.

\newpage 
The BLM and non-BLM contributions to $K_{ij}^{(2)}$ are denoted by
$K_{ij}^{(2)\beta_0}$ and $K_{ij}^{(2)\rm rem}$, respectively. The latter are
independent on $n_l$ (the number of massless quark flavors), while the former
are proportional to $\beta_0 = 11 - \f{2}{3} (n_l + 2)$. In practice $n_l=3$
because masses of the light $q=u,d,s$ quarks are neglected in loops on the
gluon lines in $b \to s\gamma$ and $b \to sg\gamma$, as well as for external
$q\bar q$ pairs in $b \to sg^*\gamma \to s q\bar q\gamma$. Although masses of
the $c$ and $b$ quarks are not neglected, all the quantities in
Eq.~(\ref{kexp}) are $\overline{\rm MS}$-renormalized at $\mu_b$ in the
five-flavor theory, which justifies the use of five-flavor $\beta_0$ in
$K_{ij}^{(2)\beta_0}$. Effects of non-zero values of $m_c$ and $m_b$ in loops
on the gluon lines are known from
Refs.~\cite{Asatrian:2006rq,Boughezal:2007ny} for all the $K_{ij}^{(2)}$ with
$i,j \in \{1,2,7,8\}$. No real $c\bar c$ pair production is included in $b \to
X_s^p \gamma$ by definition, while $b\bar b$ production is kinematically
forbidden anyway.

Contributions to $K_{ij}^{(2)\beta_0}$ from the $b \to s\gamma$ channel arise
for $(ij)=(17)$, $(27)$, $(77)$ and $(78)$ only. They were originally
calculated in Ref.~\cite{Bieri:2003ue}. Three- and four-body final state
contributions ($b \to sg\gamma$ and $b \to sg^*\gamma \to s q\bar q\gamma$)
for all the $i,j \in \{1,2,7\}$ cases and for $K_{78}^{(2)\beta_0}$ were
evaluated first in Ref.~\cite{Ligeti:1999ea}. Recently, $K_{88}^{(2)\beta_0}$
has been found by Ferroglia and Haisch~\cite{Ferroglia:2010xe}.  
\begin{table}[t] 
\begin{center} 
\begin{tabular}{|c|c|c|c|}\hline 
$(ij)$ & final state & original & confirmation \\ 
& multiplicity & calculation &\\\hline 
$(77)$ & 2 & \cite{Bieri:2003ue} & \cite{Melnikov:2005bx} \\\hline
$(77)$ & 3, 4 & \cite{Ligeti:1999ea} & \cite{Bieri:2003ue,Melnikov:2005bx}\\\hline 
$(78)$ & 2 & \cite{Bieri:2003ue} & \cite{Asatrian:2010rq} \\\hline
$(78)$ & 3, 4 & \cite{Ligeti:1999ea} & \cite{Asatrian:2010rq,Ferroglia:2010xe}\\\hline 
$(88)$ & 3, 4 & \cite{Ferroglia:2010xe} & this paper \\\hline 
$(17)$,$(27)$ & 2 & \cite{Bieri:2003ue} & \cite{Boughezal:2007ny} \\\hline 
$(17)$, $(27)$ & 3, 4 & \cite{Ligeti:1999ea} & this paper \\\hline 
$(11)$, $(12)$, $(22)$ & 3, 4 & \cite{Ligeti:1999ea} & this paper \\\hline 
$(18)$, $(28)$ & 3, 4 & this paper & \\\hline 
\end{tabular} 
\end{center} 
\caption{\sf Present status of $K_{ij}^{(2)\beta_0}$ calculations. \label{tab:status}} 
\end{table}

In the present paper, we provide the last two missing contributions, namely
$K_{18}^{(2)\beta_0}$ and $K_{28}^{(2)\beta_0}$. Moreover, we confirm the
results for $(ij)=(11)$, $(12)$, $(22)$, $(17)$ and $(27)$ from
Ref.~\cite{Ligeti:1999ea}, as well as for $K_{88}^{(2)\beta_0}$ from
Ref.~\cite{Ferroglia:2010xe}. Table~\ref{tab:status} summarizes the present status of
$K_{ij}^{(2)\beta_0}$ calculations.

The article is organized as follows. In Sec.~\ref{sec:28}, our evaluation of
$K_{18}^{(2)\beta_0}$ and $K_{28}^{(2)\beta_0}$ is presented.
Sec.~\ref{sec:22} is devoted to the remaining contributions that involve the
current-current operators ($Q_1$ and $Q_2$). Self-interference of the gluonic
dipole operator $Q_8$ is considered in Sec.~\ref{sec:88}. We conclude in
Sec.~\ref{sec:concl}.
\begin{figure}[t]
\begin{center}
\includegraphics[width=8cm,angle=0]{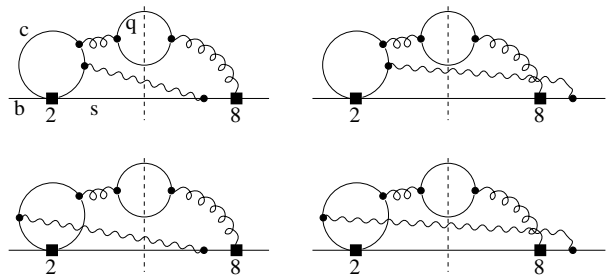}
\end{center}
\begin{center}
\caption{\sf Feynman diagrams that determine $K_{28}^{(2)\beta_0}$. \label{fig:blm28}}
\end{center}
\vspace*{-4mm}
\end{figure}

\section{Calculation of $K_{18}^{(2)\beta_0}$ and $K_{28}^{(2)\beta_0}$ \label{sec:28}}

Determination of $K_{18}^{(2)\beta_0}$ and $K_{28}^{(2)\beta_0}$ amounts to
evaluating $n_l$-dependent parts of $K_{18}^{(2)}$ and $K_{28}^{(2)}$. The
latter originates from interference of decay amplitudes generated by the
current-current operator $Q_2$ and the gluonic dipole operator $Q_8$. The
contributing Feynman diagrams are most conveniently presented using Cutkosky
rules~\cite{Cutkosky:1960sp} as four-loop propagator diagrams with unitarity
cuts. They are displayed in Fig.~\ref{fig:blm28}. In dimensional
regularization, no diagrams with cuts through the gluon lines need to be
considered because the massless $q\bar q$-loop integral is scaleless for an
on-shell gluon, which implies that all such diagrams vanish.  If $Q_2$ is
replaced by $Q_1$, the color factor gets modified according to~ $T^a \to T^b
T^a T^b = -\f16 T^a$, which leads to a simple relation
\be \label{color18}
K_{18}^{(2)\beta_0} = -\f16 K_{28}^{(2)\beta_0}.
\ee

\begin{figure}[b]
\begin{center}
\includegraphics[width=8cm,angle=0]{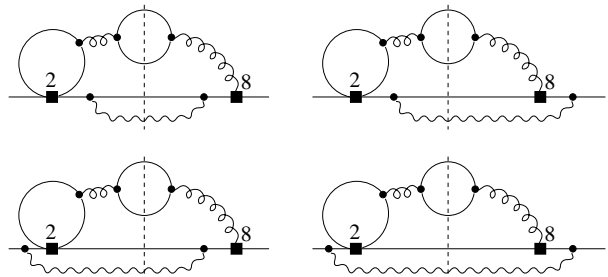}
\end{center}
\begin{center}
\caption{\sf Feynman diagrams excluded from $K_{28}^{(2)\beta_0}$. \label{fig:nonblm28}}
\end{center}
\vspace*{-4mm}
\end{figure}

Following the conventions introduced in Ref.~\cite{Ligeti:1999ea}, we exclude
the diagrams depicted in Fig.~\ref{fig:nonblm28} from the BLM approximation
despite their $n_l$-dependence. They are correlated via renormalization group
with tree-level $b \to s q\bar q\gamma$ matrix elements of the neglected
four-quark operators $Q_3$, ..., $Q_6$. Excluding those diagrams from the BLM
calculation is indeed reasonable. No other $n_l$-dependent diagrams arise
because the $Q_2$-generated charm loops vanish if the on shell photon alone
is emitted from them.

In our actual evaluation of $K_{28}^{(2)\beta_0}$, the method of Smith and
Voloshin~\cite{Smith:1994id} has been applied. It amounts to considering
lower-order diagrams that are obtained from those in Fig.~\ref{fig:blm28} by
removing the $q\bar q$ loop from the gluon propagators. However, an arbitrary
auxiliary mass of the gluon needs to be introduced. Next, integration over the
gluon mass should be performed according to the formulae of
Ref.~\cite{Smith:1994id}. 

We have carried out the calculation via direct integration over the 3-body
partly massive phase space ($m_s=0$) that is conveniently parametrized in
terms of two variables: $u = 2(p_b p_\gamma)/m_b^2$ and $s = (p_g +
p_\gamma)^2/m_b^2$.  Explicit results from Sec.~4 of Ref.~\cite{Buras:2002tp}
for the one-loop $Q_2$ amplitude with an external off-shell gluon have
appeared to be useful. Once the Dirac algebra is performed, we are left with
precisely the same two Feynman parameter integrals as in Eqs.~(4.21) and
(4.22) of that paper, namely
\bea
F_b(s,z,v) &=& \int_0^1 dx \int_0^1 dy\; \f{1}{v+ys-\f{z-i\varepsilon}{x(1-x)}},\nnb\\
F_g(s,z,v) &=& \int_0^1 dx \int_0^1 dy\; \f{-xy}{v+ys-\f{z-i\varepsilon}{x(1-x)}},
\eea
where $z = m_c^2/m_b^2$ and $v=m_{\rm gluon}^2/m_b^2$. Considering them here
in $D=4$ dimensions is sufficient because the calculation is free of
ultraviolet, infrared or collinear divergences. Integrations over the two
Feynman parameters $x$ and $y$ and over the phase-space variable $u$ are
performed analytically in a straightforward manner.  The remaining two
integrations (over $s$ and $v$) have been completed numerically. More details
will be presented elsewhere~\cite{Poradzinski:2011xxx}.

For $K_{ij}^{(2)\beta_0}$ with $i,j \neq 7$ we shall use the following
notation (consistent with Ref.~\cite{Misiak:2006ab}):
\be \label{kij}
K_{ij}^{(2)\beta_0} = 2(1+\delta_{ij}) \beta_0 \left[ \phi^{(1)}_{ij}(\delta)L_b + h^{(2)}_{ij}(\delta)\right],
\ee
where~ $\delta = 1 - 2E_0/m_b$,~ $L_b = \ln\left(\mu_b^2/m_b^2\right)$,~ and~
$\phi^{(1)}_{ij}(\delta)$\linebreak are the well-known NLO bremsstrahlung functions
collected in Appendix~E of Ref.~\cite{Gambino:2001ew}.

Our final result for the function $h^{(2)}_{28}(\delta)$ reads  
\bea \label{h28}
h^{(2)}_{28}(\delta)  
&=&     0.02605 +   0.1679 \,\delta -   0.1970 \,\delta^2            \nnb\\
&+& (  -0.03801 +   0.6017 \,\delta -   0.7558 \,\delta^2 )\, z^\f12 \nnb\\
&+& (   2.755   -  10.03   \,\delta +  11.27   \,\delta^2 )\, z      \nnb\\
&+& ( -27.05    +  68.47   \,\delta -  72.51   \,\delta^2 )\, z^\f32 \nnb\\
&+& (  85.87    - 289.3    \,\delta + 297.7    \,\delta^2 )\, z^2    \nnb\\
&+& ( -91.53    + 399.8    \,\delta - 399.9    \,\delta^2 )\, z^\f52.
\eea
The above expression is a numerical fit that remains accurate in the ranges~
$0 \leq z \leq 0.13$~ and~ $0.2 \leq \delta \leq 0.6$. These ranges will also
be valid for the fits in Sec~\ref{sec:22}. The central values
used in Eq.~(\ref{bsgsm}) are~ $\delta = 1 - 2 (1.6/4.68) \simeq 0.316$~ 
and~ $z = \left[m_c(1.5{\rm GeV})/4.68\right]^2 \simeq 0.0584$.

Eq.~(\ref{h28}) is the main new result of the present paper. Its numerical
effect on the branching ratio turns out to be miniscule (below
0.1\%). However, the purpose of the present calculation is not finding
sizeable effects but rather removing several minor uncertainties that had to
be taken into account in Refs.~\cite{Misiak:2006zs,Misiak:2006ab} in
estimating the $\pm 3\%$ perturbative error that was unrelated to the
$m_c$-interpolation.

\section{Other contributions from current-current operators \label{sec:22}}
Let us now consider $K_{ij}^{(2)\beta_0}$ for $(ij) \in \{(11), (12),
(22)\}$. The three Feynman diagrams to be calculated in the $(22)$ case are
obtained from the left parts of the cut diagrams in
Fig.~\ref{fig:blm28} by forming all the possible interference terms. The 
  cases $(11)$ and $(12)$ differ from $(22)$ by color factors 
(analogously to Eq.~(\ref{color18})), namely
\be
K_{22}^{(2)\beta_0} = -6 K_{12}^{(2)\beta_0} = 36 K_{11}^{(2)\beta_0}.
\ee
As before, only the diagrams with both the photon and the gluon coupled to
the charm loop are included in the BLM approximation for the $b \to sg^*\gamma
\to s q\bar q\gamma$ channel.

Using precisely the same methods as in Sec.~\ref{sec:28}, we obtain the
following numerical fit:
\bea \label{h22}
h^{(2)}_{22}(\delta) 
&=&     0.01370 +   0.3357 \,\delta - 0.08668  \,\delta^2            \nnb\\
&+& (   0.3575  +   1.825  \,\delta - 0.3743   \,\delta^2 )\, z^\f12 \nnb\\
&+& (  -2.306   -   5.800  \,\delta - 6.226    \,\delta^2 )\, z      \nnb\\
&+& (   3.449   -   0.5480 \,\delta + 17.27    \,\delta^2 )\, z^\f32.
\eea

Similarly, for the photonic dipole ($Q_7$) and the current-current operator
interferences, we find
\bea \label{h27}
h^{(2)}_{27}(\delta)  
&=&    -0.1755  -   1.455  \,\delta + 1.119    \,\delta^2            \nnb\\
&+& (   0.7260  -   7.230  \,\delta + 5.977    \,\delta^2 )\, z^\f12 \nnb\\
&+& (  13.79    + 113.7    \,\delta - 100.4    \,\delta^2 )\, z      \nnb\\
&+& (-145.1     - 307.1    \,\delta + 388.5    \,\delta^2 )\, z^\f32 \nnb\\
&+& ( 475.2     + 313.0    \,\delta - 775.8    \,\delta^2 )\, z^2    \nnb\\
&+& (-509.7     - 126.1    \,\delta + 646.2    \,\delta^2 )\, z^\f52,
\eea
together with $K_{17}^{(2)\beta_0} = -\f16 K_{27}^{(2)\beta_0}$.  However, the
two-body contribution $T$ is non-vanishing in this case, so instead of
Eq.~(\ref{kij}) one has
\bea
K_{27}^{(2)\beta_0} = T + 2 \phi^{(2)\beta_0}_{27} \equiv 
T + 2 \beta_0 \left[ \phi^{(1)}_{27} L_b + h^{(2)}_{27} \right].
\eea
Explicit formulae for $T$ can be found in Ref.~\cite{Misiak:2006ab}.

To compare our expressions in Eqs.~(\ref{h22}) and (\ref{h27}) to
  Ref.~\cite{Ligeti:1999ea}, we made use of their results provided to us in
  the form of a numerical grid~\cite{Ligeti:2007xxx}. The grid described
  contributions to the differential photon energy spectrum in the ranges~
  $0.2\, m_b \leq E_\gamma \leq m_b/2$~ and~ $0 \leq z \leq 0.13$.\linebreak The
  originally published fits in Eq.~(12) of Ref.~\cite{Ligeti:1999ea} were
  valid in more narrow ranges, especially in the case of $z$. After performing
  an accurate fit to the grid, we integrated the spectrum over $E_\gamma$ from
  $E_0$ to $m_b/2$, and then compared the outcome to our results for
  $h^{(2)}_{22}(\delta)$ and $h^{(2)}_{27}(\delta)$ in the ranges $0.2
  \leq \delta \leq 0.6$~ and $0 \leq z \leq 0.13$. A perfect agreement was
  found immediately at the first attempt to perform such a comparison,
  indicating undoubtedly that we can confirm the results of
  Ref.~\cite{Ligeti:1999ea}.

It is interesting to observe that $h^{(2)}_{22}$ affects the branching ratio
by $+1.9\%$, which remains within the assumed $\pm 3\%$ uncertainty for
all such effects in Eq.~(\ref{bsgsm}). Neither this contribution nor the
  smaller ones from $h^{(2)}_{27}$ and $h^{(2)}_{78}$ were 
  included in the analysis of Refs.~\cite{Misiak:2006zs,Misiak:2006ab}.

\section{Gluonic dipole operator self-interference \label{sec:88}}
The three Feynman diagrams that matter for $K_{88}^{(2)\beta_0}$ are obtained
from the right parts of the cut diagrams in Fig.~\ref{fig:blm28} by
forming all the possible interference terms. Contrary to the previously
discussed cases, a collinear divergence arises here, and a non-vanishing 
  mass of the $s$-quark must be retained in the line from which the photon
is emitted. In analogy to the NLO calculation of $K^{(1)}_{88}$, we shall keep
this mass whenever it can produce $\ln(m_b/m_s)$, but neglect all the power
corrections $(m_s/m_b)^n$.

Note that only $b \to sg^*\gamma \to s q\bar q\gamma$ matters for
$K_{88}^{(2)\beta_0}$, which means that no photon emission from the $q\bar q$
pair needs to be considered. Amplitudes with such an emission would not be
proportional to $n_l$ but rather weighted with the quark electric
charges. Consequently, the quarks into which the gluon fragments can be kept
massless from the outset, even if they are the $s$ quarks.

As the interfering amplitudes are tree-level here, all the phase-space
integrals and the ficticious gluon mass integral can be performed
analytically. We obtain
\bea
h_{88}^{(2)}(\delta) &=& \fm{4}{27} \left\{ \left[ 
\left(1 + \fm{1}{2}\delta\right)\delta\ln\delta - 6 \ln(1\!-\!\delta) - 2 {\rm Li}_2(1\!-\!\delta)
\right.\right. \nnb\\[2mm] &+& \left. \left. 
\fm{1}{3}\pi^2 - \fm{16}{3} \delta - \fm{5}{3} \delta^2 + \fm{1}{9} \delta^3 \right] \ln \f{m_b}{m_s}
- 2 {\rm Li}_3(\delta)
\right. \nnb\\[2mm] &+& \left. 
\left( 5\!-\!2 \ln \delta \right) \left[ {\rm Li}_2(1\!-\!\delta) - \fm{1}{6} \pi^2 \right]
- \fm{1}{12} \pi^2 \delta\, (2\!+\!\delta) 
\right. \nnb\\[2mm] &+& \left. 
\left[ \fm{1}{2} \delta + \fm{1}{4} \delta^2 - \ln(1\!-\!\delta) \right] \ln^2 \delta
+ \left( \fm{151}{18} - \fm{1}{3}\pi^2 \right) \times 
\right. \nnb\\[2mm] &\times& \left. 
\ln(1\!-\!\delta) + \left( - \fm{53}{12} - \fm{19}{12} \delta + \fm{2}{9} \delta^2 \right) \delta \ln \delta
\right. \nnb\\[2mm] &+& \left. 
\fm{787}{72} \delta + \fm{227}{72} \delta^2 - \fm{41}{72} \delta^3 \right\}. 
\eea

The corresponding contribution to the photon energy spectrum is found by
differentiating $K_{88}^{(2)\beta_0}$ with respect to $\delta$. Doing so, we
find perfect agreement with the very recent article of Ferroglia and
Haisch~\cite{Ferroglia:2010xe}. An extended discussion of collinear
divergences can be found there, which adds new elements to the previous
analyses in Refs.~\cite{Kapustin:1995fk,Benzke:2010js}. Replacing the
perturbative collinear regulator $m_s$ by a physical hadronic one can hardly
be performed in a quantitatively precise manner given our poor knowledge of
the QCD bound state properties. Fortunately, the gluonic dipole operator
self-interference undergoes significant suppression in $\bar B \to X_s \gamma$ due
to $(Q_d C_8/C_7)^2 \simeq 1/36$, as well as the relatively high photon
energy cut~ $E_0 \sim m_b/3$. Evaluation of $K_{88}^{(2)\beta_0}$ provides just
a check that no unexpected large numerical factors overcome this suppression
at the perturbative level. The overall effect of $K_{88}^{(2)\beta_0}$ on the
$b \to X^p_s\gamma$ decay width does not exceed $0.2\%$. 

\section{Conclusions \label{sec:concl}}

The NNLO QCD corrections to $b \to X^p_s \gamma$ in the BLM approximation
receive contributions from $b \to s\gamma$, $b \to sg\gamma$ and $b \to
sg^*\gamma \to s q\bar q\gamma$. The former results are well established by
now, while the $b \to sg\gamma$ BLM amplitudes vanish in dimensional
regularization.  In this article, we have calculated the last missing $b \to
sg^*\gamma \to s q\bar q\gamma$ contributions, namely $K_{18}^{(2)\beta_0}$
and $K_{28}^{(2)\beta_0}$. In addition, we have confirmed all the previously
known results for BLM corrections to the photon energy spectrum that involve
the current-current operators, as well as the recently found
$K_{88}^{(2)\beta_0}$. Numerical effects of all these quantities on the
branching ratio remain within the $\pm 3\%$ perturbative uncertainty 
estimated in Refs.~\cite{Misiak:2006zs,Misiak:2006ab}.

\begin{center} {\bf ACKNOWLEDGMENTS} \end{center}

We thank Andrea Ferroglia and Ulrich Haisch for letting us know the
results of Ref.~\cite{Ferroglia:2010xe} prior to publication,  Zoltan
Ligeti for an extended version~\cite{Ligeti:2007xxx} of the numerical results
of Ref.~\cite{Ligeti:1999ea}, and Matthias Steinhauser for his valuable
  comments on the manuscript. This work has been supported in part by the
Ministry of Science and Higher Education (Poland) as research project
N~N202~006334 (2008-11), by the EU-RTN program ``FLAVIAnet''
(MRTN-CT-2006-035482), and by the DFG through the SFB/TR~9 ``Computational
  Particle Physics''. M.P. acknowledges support from the EU-RTN program
  ``HEPTOOLS'' (MRTN-CT-2006-035505). M.M. acknowledges support from the
DFG through the ``Mercator'' guest professorship program.

\end{document}